\documentclass[11pt, a4paper]{article}
\pdfoutput=1
\usepackage{jheppub}
\usepackage{graphicx}
\usepackage{epsfig,amsmath,amsthm}
\usepackage{epstopdf}

\newcommand{\be}{\begin{equation}}
\newcommand{\ee}{\end{equation}}
\newcommand{\bea}{\begin{eqnarray}}
\newcommand{\eea}{\end{eqnarray}}
\def\bse{\begin{subequations}}
\def\ese{\end{subequations}}

\def\IZ{\relax\ifmmode\hbox{Z\kern-.4em Z}\else{Z\kern-.4em Z}\fi}

\newcommand{\non}{\nonumber \\}

\def\half{\frac{1}{2}} 

\def\del{{\partial}}

\DeclareMathOperator\arctanh{arctanh}

\def\tI{\widetilde{I}}

\def\al{\alpha} \def\bt{\beta}
  \def\eps{\epsilon}
\def\tbt{\tilde{\beta}} \def\tDelta{\tilde{\Delta}}

\def\lam{\lambda}

\def\presub{\vspace{.5cm} \noindent}

\def\bi{\begin{itemize}} \def\ei{\end{itemize}}

\def\({\left(} \def\){\right)}
\def\[{\left[} \def\]{\right]}
\def\<{\left<} \def\>{\right>}

\def\tI{\widetilde{I}}

\title{Bubble diagram through the Symmetries of Feynman Integrals method}

\author{Barak Kol  \\
{\it Racah Institute of Physics, Hebrew University, Jerusalem 91904, Israel} \\
{\tt barak.kol@mail.huji.ac.il}
}

\abstract{The Symmetries of Feynman Integrals method (SFI) associates a natural Lie group with any diagram, depending only on its topology. The group acts on parameter space and the method determines the integral's dependence within group orbits.  This paper analyzes the bubble diagram, namely the 1-loop propagator diagram, through the SFI method. This is the first diagram with external legs to be analyzed within SFI, and the method is generalized to include this case. The set of differential equation is obtained.  In order to solve it the set is transformed into partially invariants variables. The equations are integrated to reproduce the integral's value. This value is interpreted in terms of triangle geometry suggested by extant papers. }

\begin{document}
\maketitle

\section{Introduction}
 
Feynman diagrams and their associated expressions, introduced by Feynman in 1948-9 \cite{FeynDiag,WikiFeynDiag}, arguably form the backbone of Quantum Field Theory. After taking care of tensor algebra their computational core is seen to be a certain class of scalar integrals known as Feynman Integrals. While Feynman integrals may diverge, regularization and evaluation of a finite part are always possible and physically meaningful. It is standard to use dimensional regularization \cite{DimReg}. 

\begin{figure}
\centering \noindent
\includegraphics[width=8cm]{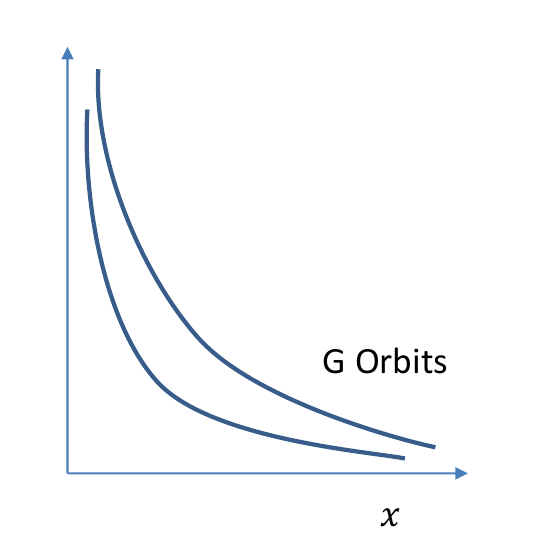}
\caption[]{This figure demonstrates schematically the main result of the method of Symmetries of Feynman Integrals (SFI). A Feynman diagram defines its parameter space denoted by $x$ and a group $G$ which acts on it. Accordingly, the $x$ space is foliated into $G$ orbits. The method defines a set of differential equations for the integral $I(x)$ within the $G$ orbits.}
 \label{fig:central}
\end{figure}

Along the years an assortment of methods were devised for evaluating Feynman Integrals, see for example \cite{SmirnovBooks} and references therein. Recently the Symmetries of Feynman Integrals (SFI) method was introduced \cite{SFI,locus} (see also developments to appear in \cite{ExactDiameter,VacuumSeagull,TriangleDiag}). In this method one starts by considering a diagram of fixed topology and considering its dependence on the widest possible set of parameters. So far the vacuum diagrams were considered and the parameters consisted of all possible masses. In this paper we shall consider diagrams with external legs where the parameter space will be supplemented by the kinematical invariants -- Lorentz scalars formed by the external momenta. The SFI method defines a set of partial differential equations for dependence of the Feynman integral on its parameters, closely related to the ones defined in \cite{Tarasov1998}, see also \cite{KalmykovKniehl2012}. This equation set defines a Lie group $G$, not to be confused with the familiar discrete symmetry group of the diagram denoted here by $\gamma$, but related to the group defined by R. Lee \cite{LeeGroup}. $G$ in turn foliates the parameter space into $G$-orbits, such that the equation set implies the dependence within orbits, see figure \ref{fig:central}. In other words, the SFI method defines the dependence of the integral on some of its parameters through differential equations.

The Symmetries of Feynman Integrals method is rather general: it makes no assumption on the space-time dimension nor on the values of the parameters. It enables to reduce the evaluation of any integral to an evaluation at a conveniently chosen parameter point on the same $G$-orbit followed by solving the equation set (which was shown to reduce itself to a line integral). The former integration must be evaluated by some other method such as integration of alpha (or Schwinger) parameters, possibly numerically. 

The SFI method is closely related to two widely used methods -- the Integration By Parts (IBP) method \cite{ChetyrkinTkachov81} and the Differential Equations (DE) method \cite{Kotikov1990,Remiddi1997,GehrmannRemiddi1999}, see also references within \cite{HennRev2014}. In a sense SFI unifies the two, or at least stresses their unity, by showing that the recursion relations of IBP and the differential equations of DE are related to each other by a transformation of the independent variables. 

Recent work on IBP includes  an expression for the number of master integrals \cite{LeePomeransky2013};  \cite{KniehlTarasov2016} where master integrals are counted and functional equations are illustrated;  a determination of the kite integral \cite{Adams:2016xah}; a connection with unitarity cuts and syzygy equations \cite{LarsenZhang2015}; \cite{Forcer,Hoff:2016pot,Gituliar:2016vfa} describing new IBP related programs; \cite{Kompaniets:2016hct} considering renormalization group functions; \cite{Schnetz:2016fhy} studying numbers and functions which appear in QFT;  and finally \cite{Eden:2016dir,Henn:2016men,Marquard:2016dcn} which study certain 4-loop integrals.

In \cite{SFI} the Symmetries of Feynman Integrals method was demonstrated by application to the two-loop vacuum diagram. In this paper we apply it to the bubble diagram, see fig. \ref{fig:bubble}. The value of this integral is known in full generality and can be obtained through direct integration of the alpha (or Schwinger) parameters. We consider this diagram since it requires to formulate the method for diagrams including external legs. Other results shall follow as described below.

The structure of this paper is as follows. We start in section \ref{sec:eq} by formulating the SFI equation set in the presence of external legs and finding a relation with the diagram's vacuum closure. In section \ref{sec:bubble} we turn from the general procedure to the specific case of the bubble diagram and we obtain the SFI equation set, as well as the structure and orbits of the associated group $G$. In section \ref{sec:soln} we proceed to solve the equations. We find the homogeneous solution, the algebraic locus \cite{locus} and the solution there. Then the equation set is transformed into partially invariant variables and integrated. In section \ref{sec:triangle} we discuss a geometrical interpretation of the solution in terms of triangle geometry found in \cite{DavydychevDelbourgo1997}. Finally we summarize our results and add a discussion in section \ref{sec:summ}. Some extra material appears in the appendices.

{\bf v3} (September 2018): This version corrects the description of the singular locus -- see in section  \ref{sec:soln} the ``singular locus'' part. The reason for the mistake was an erroneous use of the diagram's reflection symmetry. The new text provides the correct singular locus together with the diagram's value on it, using methods that were developed in the meantime. In addition,  a useful covariant basis for the SFI equations is given in section \ref{sec:bubble}.

\section{Equation set with external legs}
\label{sec:eq}

In this section we extend the SFI equation set to include external legs. In the next section we shall use this formulation to treat a specific case, that of the bubble diagram.

Consider a general Feynman diagram with $L$ loops and $n$ external legs. We associate with it a rather general Feynman integral \be
 I(\mu,\, p) =  \int \frac{dl}{\prod_{i=1}^P \(k_i^2-\mu_i + i 0\)} ~,
\label{def:I}
 \ee
where each one of the $P$ propagator lines is associated with a mass-squared parameter $\mu_i \equiv m_i^2, ~ i=1,\dots, P$ $p_u, ~u=1,\dots,n-1$ are a choice of $n-1$ independent external currents (currents starting at infinity and ending there); the integration measure is over some choice of $L$ loop current variables in $d$ space-time dimensions, namely $dl := \Pi_{r=1}^L d^d l_r$  and the propagator currents are expressed as a linear combination of  loop and external currents \be
 k_i = A_i^{~r}\, l_r + B_i^{~u}\, p_u ~,
 \label{def:AirBiu}
\ee
 where $A_i^{~r},\, B_i^{~u}$ are the components of the linear transformation. Due to Lorentz invariance the integral can depend on the external momenta only through their scalars $p_u \cdot p_v$ hence the integral's parameters space is described by \bea
 I &=& I(x) \non
 \mbox{where} && \non
   \{ x \} &:=& \left\{ \mu_i,\,  p_u \cdot p_v \right\}    ~. \eea
 $I(x)$ is also a generating function for general indices \cite{SFI}. As usual the definition is independent of the choice of loop and external currents and in particular the integral measure can be represented more symmetrically as an integral over all propagator currents multiplied by delta functions which enforce current conservation at all vertices.

The SFI method associates with a diagram a set of differential equations for the Feynman integral and a related Lie group $G$ \cite{SFI}. First we recall how this is done for a vacuum diagram \cite{SFI}.

\presub {\bf Vacuum diagram}. For a vacuum diagram the group $G$ associated with the SFI equation set can be defined through current freedom  \cite{locus,VacuumSeagull} as follows. One is free to choose any set of loop currents as long as the two sets are related by an invertible linear transformation, namely an element of $GL(L)$. In order to obtain the SFI equation set one considers infinitesimal changes of currents \be
 \delta l_r = \(T_C\)_r^{~s}\, l_s
 \label{def:vac-generator}
\ee
where the generator $ \(T_C\)_r^{~s}$ is a real valued matrix and $C$ stands for currents. One defines the space of squared propagator currents by \be
 S :=  Sp\{k_i^2 \}_{i=1}^{P}
 \ee
 namely the space spanned by all the squares of propagator (or edge) currents, where $P$ is the number of propagators in the diagram. Since the propagator currents are linear combinations of loops currents, any generator  (\ref{def:vac-generator})  induces a variation of the squared propagator currents and hence of $S$. The group $G$ is defined to be the subgroup of $GL(L)$ which preserves $S$, namely \be
  G \subset GL(L) \mbox{ is defined to preserve } S
 \label{preserve-vac}
 \ee
 In IBP language this precisely means that no numerators are generated by the generators of $G$, namely $G$ is numerator free.
 
Operating with any generator in $T^A \in G$ of the form (\ref{def:vac-generator}) on the integral (\ref{def:I}) a set of differential equations is generated \be
	0 =   c^A\, I  + \(T_\mu^{~A}\)^i_{~j}\, \mu_i \, \frac{\del}{\del \mu_j}\,  I + J^A   ~.
\label{SFImu}
\ee
where $T_\mu$ is the induced representation of $G$ on the $\mu$ space whose form is detailed in appendix \ref{app:induced}.

\presub {\bf Including external legs}. The loop current variation (\ref{def:vac-generator}) can now be generalized to depend on $p$ as well \be
 \delta l_r = \(T_C\)_r^{~s}\, l_s + T_r^{~u}\, p_u ~.
 \label{var-loop}
\ee
We further consider changes in the external currents of the form \be
 \delta p_u = \(T_C\)_u^{~v}\, p_v  ~. 
 \label{var-ext}
\ee
This kind of variation would not leave $I(\mu,\, p)$ invariant, but by equating the action on the integrand with the action on the integral we may still get useful equations.  However, a variation of the form $\delta p \simeq l$ does not make sense because $l$ is an integration variable, while $p$ is not.

Summarizing the previous paragraph we can consider current variations of the following form
\be
\delta \left[ \begin{array}{c}
 l_r  \\
 p_u
\end{array} \right] =
\left[ \begin{array} {cc}
\(T_C\)_r^{~s} 	& 	\(T_C\)_r^{~v} \\
0 			&	\( T_C\)_u^{~v} 
\end{array} \right] \, 
 \left[ \begin{array}{c}
 l_s  \\
 p_v
\end{array} \right]
\label{variation}
\ee
where the generator matrices $T_C$ are real valued. Said differently $G \subset T_{L,n-1}$ where $T_{L,n-1}$ are the block upper triangular matrices such that the first block is of size $L$ and the second one is $n-1$. 

For $T_C$ to generate a differential equation (and belong to $G$) we still need to generalize the condition (\ref{preserve-vac}). Requiring that no numerators are generated from $k_i^2$ means that $T_C(S) \subset S \oplus Q_p$ where $Q_p$ the space of quadratics in external currents is defined by \be
Q_p := Sp\{p_u \cdot p_v \}_{u,v=1}^{n-1}
\label{def:Qp}
\ee
Since (\ref{var-ext}) implies that $T_C(Q_p) \subset Q_p$ we see that the generalized condition for $G$ is  \be
  G \subset T_{L,n-1} \mbox{ is defined to preserve } S \oplus Q_p ~. 
 \label{preserve}
 \ee

A set of differential equations is defined in analogy with the vacuum case (\ref{SFImu}) \be
	0 =   c^A\, I  + \(T_X^{~A}\)^i_{~j}\, x_i \, \frac{\del}{\del x_j}\,  I + J^A   ~.
\label{SFIx}
\ee
 only here $T_X$ is the induced representation of $G$ on the whole $x$ parameter space gotten by operating with $T^A \in G$ of the form (\ref{variation}) on the integral  (\ref{def:I}) 

\presub  {\bf Comments}. 

Loop subgroup. One can define the loop subgroup of G \be
 G_l := \mbox{variations of the form (\ref{variation}) such that } \( T_C\)_u^{~v}=0
\label{def:Gl}
\ee
These are generators which vary loop currents but not external currents. They form a normal subgroup of $G$, namely \be
G_l \triangleleft G ~.
\label{Gl-normal}
\ee
For $T \in G_l$ the associated differential equation includes $\del/ \del \mu$ but not $\del/\del\(p^2\)$, since the latter are generated only when one varies external currents.

Kinematics-only differential equations. Often one is interested in an integral with fixed mass parameters and variable kinematical invariants, since the participating particles and their masses are fixed by the experimental setup. For that purpose the SFI equation set (\ref{SFIx}) can be combined to generate a system of differential equations where only derivatives with respect to the kinematical invariants $\del/\del\(p^2\)$ appear, but not derivatives with respect to the square masses (this is exactly opposite to the loop subgroup). Such equations are gotten by a linear combination of (\ref{SFIx}) equations with $x$ dependent coefficients. Hence the resulting set is generally not linear in $x$ and commutators do not close.

Relation with IBP. In standard usage of the Integration By Parts (IBP) method, one considers integrals with arbitrary indices (powers of propagators) and all variations of the form (\ref{var-loop}) lead to recurrence relations.  \eqref{var-ext} is usually considered to lead to a differential equation (actually it leads to a mixed differential -- recursion  equation). Alternatively, \cite{BaikovSmirnov2000} showed that it could be considered to lead to IBP-like recurrence relations once $I$ is Taylor expanded in the kinematical invariants. From the SFI perspective both types of variations lead to differential equations which together form the SFI equation set, and hence they are treated on the same footing. Moreover, while the variations of the form (\ref{var-loop}) can be interpreted in terms of the elementary method of integration by parts, this does not seem to be the case for \eqref{var-ext}. Moreover, the term IBP fails to reflect the method's dependence on the diagram topology. For these reasons I believe that  the term ``IBP'' is not an optimal name for this method.

A mathematical perspective. We found that $G$ is defined to preserve both $Q_p$ and $S \oplus Q_p$. In linear algebra a list of increasing subspaces $\{0\} \subset V_1 \subset V_2 \subset \dots \subset V$ is called \emph{a flag}, and here $V_1 = Q_p,\, V_2 = S \oplus Q_p$. A transformation which preserves the flag, namely its stabilizer is called a parabolic subgroup \cite{WikiFlag}. In algebraic geometry a parabolic subgroup $P \subset G$ are characterized by the condition that G/P is a ``complete variety'', where the latter is an analogue of compactness in algebraic geometry \cite{WikiBorelSub,WikiCompleteVar}. 
 
\presub {\bf Vacuum closure}. Motivated by \cite{BaikovSmirnov2000,Baikov1996} we considered the vacuum diagram associated with the given diagram by adding a point at infinity where all the external legs are attached. We refer to it as the diagram's vacuum closure. 

The vacuum closure diagram maintains the same currents structure as the original diagram, only the external currents become loop currents. Let us denote by $S_{vac}, G_{vac}$ the space of the squared propagator currents in the vacuum closure and the associated group, respectively. Since the space of loop currents can only increase as we pass to the vacuum closure, we have $S_{vac} \supset S$, and hence any generator which preserves $S_{vac}$ also preserves $S$. Therefore \be
 G \supseteq G_{vac} \cap T_{L,n-1}
 \label{Gvac}
\ee
Furthermore, for $n=2,3$ external legs, $Q_p \simeq S_{vac}/S$ as seen by a dimension counting argument: on the one hand $\mbox{dim}(Q_p)=n(n-1)/2$ while on the other $\mbox{dim}(S_{vac}/S)=n$ for $n \ge 3$ and $\mbox{dim}(S_{vac}/S)=1$ for $n =2$, so for $n=2,3$ we have equality of dimensions (this is related to the fact that a triangle is defined by its side lengths). Hence for $n=2,3$ external legs variations which preserve $S_{vac}$ also preserve $Q_p \oplus S$ and the inclusion (\ref{Gvac}) is saturated into an equally. 

Comment: in comparing with \cite{BaikovSmirnov2000}  we find that (\ref{variation}) is consistent with the type of generators mentioned in that derivation, but not with the announced result regarding equivalence with the vacuum closure.

\newpage
\section{The bubble diagram}
\label{sec:bubble}

\begin{figure}
\centering \noindent
\includegraphics[width=10cm]{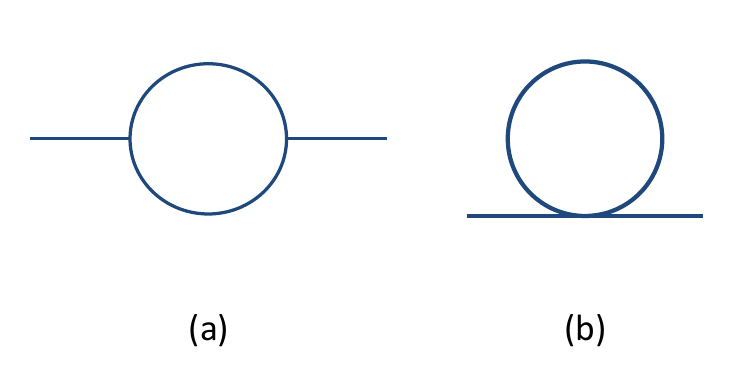}
\caption[]{(a) The 1-loop propagator diagram which we shall call the bubble diagram. (b) A diagram defining the source function $j_i$ (\ref{def:ji}).}
 \label{fig:bubble}
\end{figure}

From this section onward we consider the bubble diagram shown in fig. \ref{fig:bubble}(a). The associated integral is \be
I(p^2;\, \mu_1,\mu_2 ) = \int \frac{d^d l} { \(k_1^2-\mu_1 \)\,   \(k_2^2-\mu_2 \)} = \int \frac{d^d l} { \( (p/2+l)^2-\mu_1 \)\,   \( (p/2-l)^2-\mu_2 \)}
\label{bubble-integ}
\ee
The parameter space is composed of the two masses-squared $\mu_1,\mu_2$ and of $p^2$, the single kinematical invariant.

The diagram has a discrete reflection symmetry with respect to a horizontal axis acting by $l \to -l$ and $p \to p$, namely $\gamma = \IZ_2$. This symmetry exchanges the masses $\mu_1 \longleftrightarrow  \mu_2$. Another reflection, this time with respect to a vertical axis, acts by $p \to -p$ but it does not act on parameter space, and hence will not concern us.

The \emph{Euclidean domain} will refer to the region in parameter space where  \be
 \mu_1,\mu_2,-p^2 \equiv p_E^2 \ge 0 ~,
 \label{def:Euclid}
 \ee
  because after analytic continuation to Euclidean space-time $-k^2 \to k_E^2$ for all the momenta including $p$, the bubble's integrand (\ref{bubble-integ})
 becomes positive everywhere, and the integral will have no divergences in the interior of the domain. It is shown in fig. \ref{fig:xspace} as part of the projective parameter space.
 
From (\ref{Gvac}) we have \be
 G = T_{1,1} \subset GL(2) ~,
\label{G}
 \ee
 namely the associated group is a the 3 dimensional group of upper triangular $2 \times 2$ matrices. These are of the form  \be
  G= \left[ 
  \begin{array}{cc}
  * & * \\
  0 & * 
  \end{array}
  \right] 
 \ee
 where a star denotes an unconstrained entry.

Performing the variations (\ref{variation}) the SFI equation set is found to be \be
0 = \left[ \begin{array}{c}
2(d-3) \\
0 \\
-2 \end{array} \right] I - \[ \begin{array}{ccc}
 3 \mu_1 + \mu_2-p^2 	&	\mu_1 + 3 \mu_2 - p^2 	& 0 	\\
 \mu_1 - \mu_2 + p^2 	&	\mu_1 - \mu_2 - p^2 		& 0 	\\
\mu_1 - \mu_2 + p^2		& -\mu_1 + \mu_2 + p^2		& 4 p^2 
\end{array}\] \, \left[ \begin{array}{c}
	\del_1\, I \\
	\del_2\, I \\
	\del_{p^2}\, I
\end{array} \right]
 + \left[ \begin{array}{c}
	-j_1'-j_2' \\
	j_1'-j_2' \\
	j_1' + j_2' 
\end{array} \right]  \(  \begin{array}{c}
  \ref{SFIbubble} a \\
  \ref{SFIbubble} b \\
  \ref{SFIbubble} c 
  \end{array}\) 
\label{SFIbubble}
\ee
 where the source functions $j_i$ which appear as sources are given by the Feynman integral associated with the tadpole diagram  fig. \ref{fig:bubble}(b) 
 \be
j_i :=I_{fig. \ref{fig:bubble}(b)}(x_i) 
\label{def:ji}
\ee and hence \be
 j '(\mu) =  i \pi^{d/2}\, \Gamma\(2-\frac{d}{2}\)\, \mu^{\frac{d}{2}-2} ~,
\label{j_value}
\ee 
see appendix \ref{app:Schwinger} for the derivation through integration of the alpha parameter. The 3 equations are associated with the generators $A := 2 l\, \del_l,\, B: = p\, \del_l$ and $C := 2 p \, \del_p$ respectively.
 
It is recommended to choose the generators which appear in the equations set such that they are compatible with the discrete symmetry $\gamma$, namely belonging to particular representations of $\gamma$. Indeed the generators in (\ref{SFIbubble}) were chosen in such a way: $A$ and $C$ are even with respect to $\gamma$ while $B$ is odd. 

\presub {\bf Group structure}.  The group structure of $G=T_{1,1}$ is given by the following commutation relations $ \[ A,C \] =0, \, \[A,B\] = -2B$ and $\[B,C \] = -2 B$. Hence the derived group is $G^{(1)} = Sp\{ B \}$. 

\presub {\bf Covariant basis for SFI equation system}. Following \cite{kite} we present the SFI equation system also in a different basis, the covariant one, where it takes the following form \be
0 = \left[ \begin{array}{c}
 d-3 \\
d-3 \\
d-4 \end{array} \right] I - 2 \[ \begin{array}{ccc}
 s^3 		&	~ \mu_2	& ~0 	\\
 \mu_1 	&	~ s^3 		& ~ 0 	\\
\mu_1 	& ~ \mu_2 		& ~ p^2 \\ 
\end{array}\] \, \left[ \begin{array}{c}
	\del_1\, I \\
	\del_2\, I \\
	\del_{p^2}\, I
\end{array} \right]
 - \left[ \begin{array}{c}
	j_1' \\
	j_2' \\
	0
\end{array} \right]  ~.
\label{cov_SFI_basis}
\ee
Here we defined \be
 s^3 := (\mu_1+\mu_2-p^2)/2 \ee
 following \cite{diameter}.

This basis is covariant in the sense that the second equation can be gotten by transforming the first one under the reflection symmetry $\gamma$, namely $1 \leftrightarrow 2$ . The last equation is the standard dimension equation and is a singlet under $\gamma$.

This basis is related to the previous $A, B, C$ basis as follows \be
\left[ \begin{array}{c}
 E^1 \\
 E^2 \\
 E^3 \end{array} \right] = \left[ \begin{array}{c}
 \half (A-B) \\
\half (A+B)  \\
\half (A+C )\end{array} \right] \equiv
 \left[ \begin{array}{c}
 (l - \half p ) \del_l  \\
 (l + \half p ) \del_l \\
 l \del_l + p \del_p \end{array} \right] 
\ee


\presub {\bf Group orbits}. Inspecting the group generators in (\ref{SFIbubble}) one finds that at a generic point $x$ in parameter space the 3 vectors implied by the generators $\(T_X^{~A}\)^i_{~j}\, x_i \, \del/\del x_j, ~ A=1,2,3$ are linearly independent. This means that the generic $G$-orbit is 3 dimensional, or that its co-dimension is 0. Hence by solving the SFI equations one could obtain the dependence of the integral throughout parameter space, possibly up to a discrete number of base points (later we shall see that regularity would supplies the necessary boundary conditions thereby obviating base points).

\presub {\bf Kinematics-only differential equation}. Following the procedure described in the comment below (\ref{Gl-normal}) we obtain a single kinematics-only equation \bea
 0 &=&  \[ -(\frac{d}{2}-1)\, (\mu_1-\mu_2)^2 +  (\mu_1 + \mu_2)\, p^2 + (\frac{d}{2}-2)\, p^4 \] I -  p^2 \lambda \frac{\del}{\del p^2} I + \non
 &+& j_1'\, \mu_1 (\mu_1-\mu_2-p^2) + j_2'\, \mu_2 (\mu_2-\mu_1-p^2)  ~,
 \label{KObubble}
 \eea
where $\lam$ is defined and discussed later in \eqref{def:lam}. Upon setting $m_2=0$ this equation was tested to imply a differential equation for $\del I/\del (m^2)$ which appears in \cite{SmirnovBooks} eq. (1.23). 

The loop subgroup is generated by the $A$ and $B$ generators.

\section{Solution}
\label{sec:soln}

In this section we solve the equation set (\ref{SFIbubble}). We start by finding the homogeneous solution.

\presub {\bf Homogeneous solution}. A procedure involving invariants of the constant-free subgroup was described in \cite{locus} starting above eq. (3.7). Here the procedure can be simplified a bit by a linear re-definition of $I$ according to  \be
 I = \frac{1}{\sqrt{-p^2}}\, I'
 \label{def:Iprime}
 \ee
The re-definition is designed such that (\ref{SFIbubble}c) becomes constant free. Here and below the expression is chosen to be presented as depending on $-p^2$, rather than on $p^2$ in order to render the behavior in the Euclidean domain (\ref{def:Euclid}) more transparent. 

Now one seeks an invariant of the constant-free subgroup which consists of  (\ref{SFIbubble}b,c). Following the procedure the invariant is found to be \be
\mu_0 := \frac{\lambda}{-4 p^2}
\label{def:mu0}
\ee
where 
 \be
\lam :=  \mu_1^2 + \mu_2^2 + p^4 - 2 \mu_1\, p^2 -2 \mu_2\, p^2  -2 \mu_1\, \mu_2  ~.
\label{def:lam}
 \ee
$\lam=\lam(\mu_1, \mu_2, p^2)$ is the K\"all\'en function (see more in \cite{locus} eq. (4.8)) and it would be further interpreted around \eqref{Heron}. In that section we would also explain the notation $\mu_0$ and the normalization constant $4$.

Using (\ref{SFIbubble}a) the homogeneous solution $I_0$ can now be determined to be proportional to \be 
I_0 \propto \frac{1}{\sqrt{-p^2}}\, \( \frac{\lam}{-4 p^2}\)^{\frac{d-3}{2}} ~.
\ee
This expression is seen to be proportional to an expression involving the function $j'(\mu_0)$ which we take to define the homogeneous solution $I_0$
\be
 I_0 =  \frac{\sqrt{\lam}}{-2 p^2} ~~ j'\(\frac{\lam}{-4 p^2} \)
\label{I0}
\ee

A somewhat different way to reach the same result would be to consider invariants under the generator of (\ref{SFIbubble}b), the one which is constant-free before changing to $I'$. The space of invariants is found to be generated by $\lam,\, p^2$. Then one seeks $I_0=I_0(\lam,\, p^2)$ which satisfies the other two equations and reproduces (\ref{I0}).


\presub {\bf Singular locus}. The algebraic locus \cite{locus} is the singular locus in parameter space where the set of differential equations (\ref{SFIbubble}) degenerates into an algebraic equation for $I$ in terms of the sources. This property is related to the IBP property that the master integrals do not include the integral with all the indices equal to unity.

\begin{figure}
\centering \noindent
\includegraphics[width=8cm]{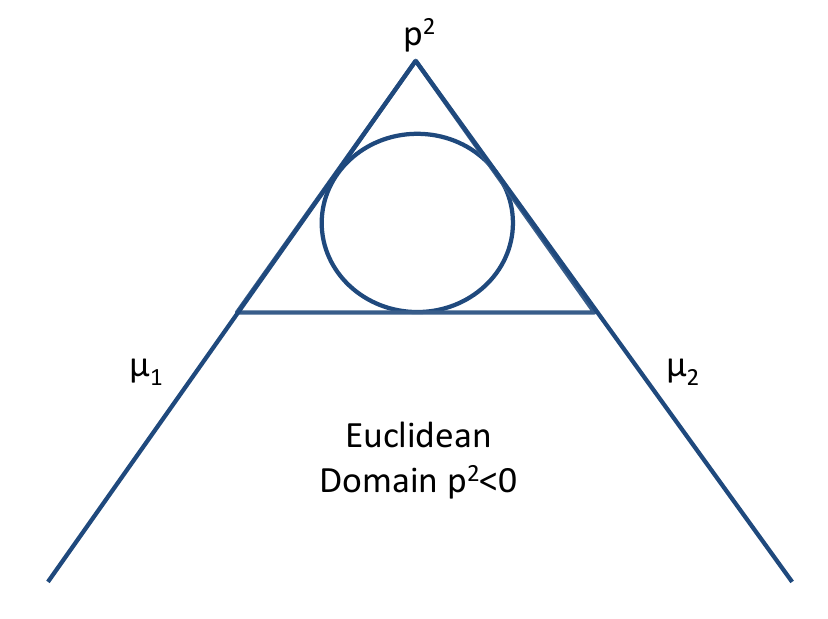}
\caption[]{The parameter space for the bubble diagram. The problem has 3 parameters $\mu_1,\mu_2,p^2$. To eliminate scale, a cross section in shown through $\mu_1 + \mu_2 + p^2 = const$. The triangle is the locus where at least one of the coordinates vanishes, the label next to a vertex means it is the only coordinate not vanishing there, and finally the circle is where $\lam$ (\ref{def:lam}) vanishes. The singular locus is composed of the $\lam=0$ circle and the $p^2=0$ horizontal line.}
 \label{fig:xspace}
\end{figure}

To find the singular locus according to the method of maximal minors \cite{minors} we compute the determinant of $Tx$ which is defined to be the $3*3$ matrix which appears in (\ref{cov_SFI_basis}) and it is found to be proportional to \be
S = p^2 \cdot \lambda ~. 
\label{sing}
\ee
$S$ is known as the singular factor and its zeros define the singular locus. Its form implies that this locus is made out of two components, one at $\lambda=0$ and the other at $p^2=0$ -- see fig. \ref{fig:xspace}. Now we turn to study these components.

$\lam=0$. Physically, this locus describes a threshold where $p= |m_1 \pm m_2 |$, where a positive sign is known as a real threshold and a negative sign is known as a pseudo-threshold. At this locus the stabilizer is given by either of the following forms \bea
Stb_{\lam 1} &=& \begin{pmatrix} \mu_1, & \; -s^3, & \;0 \end{pmatrix} \non
Stb_{\lam 2}&=& \begin{pmatrix} -s^3 , & \; \mu_2, & \;0 \end{pmatrix}
\eea
They are obtained from the $2*2$ block of $Tx$. At the $\lam$ locus the two forms are parallel.

The algebraic solution is given by \be
\left. I \right|_{\lam} =  \frac{c_b}{(d-3)\, p^{d-2}} \[ (s^1)^{d-3} + (s^2)^{d-3} \] 
\label{I_lam}
\ee  
where $c_b$, the bubble constant, is given by \be
 c_b = i \pi^{d/2}\, \Gamma\(2-\frac{d}{2}\) ~.
\ee
This form of the algebraic solution can be gotten by using the stabilizer $Stb_{\lam 1}+Stb_{\lam 2}$. 

$p^2=0$. Physically, this locus describes a degeneration into a vacuum (tadpole-like) diagram. At this  locus the stabilizer is given by \be
Stb_{p^2} = \begin{pmatrix} 2 \mu_1, & \; -2 \mu_2, & \; \mu_2-\mu_1 \end{pmatrix}
\ee
This result can be achieve through computation of 2-minors after factoring out $\mu_2-\mu_1$.

The associated algebraic solution is given by \be
\left. I \right|_{p^2} = \frac{j(\mu_2) - j(\mu_1)}{\mu_2-\mu_1}
\label{I_p2}
\ee
In deriving this one uses the relation $\mu\, j'(\mu) = \frac{d-2}{2} j(\mu)$. The same expression can be gotten from the perspective of a vacuum diagram by considering the definition of the integral (\ref{bubble-integ}) and using a decomposition into partial fractions. 


\presub {\bf Change of variables into partial invariants}.  We found that the equation set appears simplified after changing variables \be
  \mu_1,\mu_2, p^2 \to \Delta, \lam, p^2
\label{def:partial_invariants}
\ee 
 where $\Delta$ is defined by \be
\Delta := \mu_1-\mu_2 =0
\label{def:Delta}
\ee
 and $\lam$ was defined in (\ref{def:lam}).
 These variables are motivated by their appearance above. More specifically $\lam$ is invariant under the trace-free subgroup of $G$, while $p^2$ is invariant under the derived group $G^{(1)} \equiv \[ G,G \]$ generated by $p\, \del_l$ (generator B). Therefore we refer to these variables as \emph{partial invariants}. Finally $\Delta$ appears in the expression for the algebraic locus.

The transformed equation set  becomes \begin{subequations} \begin{align}
 0 =& -2 p^2 \del_\Delta I + j_1'-j_2'      \label{transformed-set_a} \\ 
 0 =&  (d-2)\, I + 2 p^2 \del_{p^2} I - j_1'-j_2'      \label{transformed-set_b}\\ 
 0 =& (d-4)\, I - \( 2 \Delta\, \del_\Delta + 4 \lam\, \del_\lam + 2 p^2\, \del_{p^2} \) I   \label{transformed-set_c}
 \end{align}
  \label{transformed-set}
\end{subequations}
where the equations correspond to the following generators $p\, \del_l, \, l\, \del_l - p\, \del_p,$ and $l\, \del_l + p\, \del_p$.

\subsection*{Integration}

We proceed to integrate the transformed equation set (\ref{transformed-set}). We choose to start by integrating \eqref{transformed-set_c}
which corresponds to integration along flow lines of $G^{(1)}$. Since the tadpole sources $j_i\ \equiv  j(\mu_i)$ are functions of $\mu_i$ we need to invert the transformation (\ref{def:partial_invariants}) and we find \be
\mu_{1,2} = \frac{1}{4 p^2} \[ \(\Delta \pm p^2 \)^2 - \lam \]
\label{mus}
\ee
where the plus sign corresponds to $\mu_1$. 

The resulting integral is of the form \be
 \int^\Delta d\Delta'\,  Q(\Delta')^{a-1}
 \ee
 where $Q(\Delta')$ is some quadratic function, and $a$ is a constant. After a linear change of variables \be
 \tDelta := \frac{2\Delta -\Delta_1 - \Delta_2}{\Delta_2-\Delta_1}
 \label{def:tDelta}
 \ee
 where $\Delta_1 \le \Delta_2$ are the roots of $Q$,
 the integral is found to be 
\be
I = \frac{\sqrt{\lam}}{-2 p^2} ~~ j'\(\frac{\lam}{-4 p^2} \) \[ B\(\tDelta_2; d/2-1 \) - B\(\tDelta_1; d/2-1  \) + C\(p^2, \lam\) \]
\label{int1}
\ee 
where the function $B(x;a)$ of a variable $x$ and a parameter $a$ is defined to be \be
 B(x;a) :=  \int_0^x (1-t^2)^{a-1} \, dt 
 \label{def:B}
 \ee
 and will be discussed immediately below, where  $\tDelta_i, ~i=1,2$ are defined by \be
 \tDelta_{1,2} := \frac{\Delta \pm p^2}{\sqrt{\lam}}
\label{def:tDeltai}
\ee
and where $C\(p^2, \lam\)$ is the integration constant. In reaching \eqref{int1} we have also used the power-law nature of $j$ which implies that $j(k\, \cdot \mu) = k^{d/2-2} \cdot j(\mu)$.

The function $B(x;a)$ defined in (\ref{def:B}) is odd $B(-x;a)=-B(x;a)$ and is well defined for $|x| <1$ (it can be analytically continued to the whole complex plane). It can be expressed in terms of the a special kind of the incomplete beta function $B(x; a,b)$ where the $a, b$ parameters are equal \be
 B(x;a) = 2^{2 a-1}\, \[ B(\frac{x+1}{2};a,a)-B(\half; a,a) \]~.
 \ee
 The incomplete beta function, in turn,  is a special case of the hypergeometric function ${}_2 F_1$ \be
 B(x; a,b) = \frac{x^a}{a}\,  {}_2 F_1 (a,1-b;\, a+1;\, x) ~.
 \ee
We note that for integral values of $d$  $B(x;a)$ can be expressed as an integral over a rational function. In fact this can be done in two different ways. In the first, motivated by the geometrical interpretation to be discussed in the next section and by \cite{DavydychevDelbourgo1997} eq. (4.8) one substitutes  $t=\tanh w$ to obtain \be
B(x;\, a) =  \int_0^{\arctanh x} \frac{1}{\cosh^{2a} w} \, dw
\ee
 where $2a = d-2$ and the integrand is rational after transforming into $e^w$. A second possibility is to substitute  $t=\sin \theta$ to get \be
B(x;\, a) =  \int_0^{\arcsin x} \cos^{2a-1} (\theta)\, d\theta
\ee
where $2a-1=d-3$, and the integrand is rational after transforming into $e^{i \theta}$.  Finally, expanding around $d=4$, namely taking $d=4-2 \eps$ we have \be
B(x; a) =   x - \eps \[ (1+x) \log (1+x) - (1-x) \log (1-x) \] + \dots 
\ee

In the Euclidean domain (\ref{def:Euclid}) $\tDelta_i$ are in the range $-1 \le \tDelta_i \le 1$ such that function $B(x;a)$ is well defined. Outside this domain analytic continuation may be required.
 
Substituting (\ref{int1}) into (\ref{transformed-set_b},\ref{transformed-set_c}) we find that $\del_{p^2} C = \del_\lam C =0$ and hence \be
C\(p^2, \lam \) = C ~.
\label{int2}
\ee

This residual freedom in the solution amounts to the possibility of adding the homogeneous solution $I_0$ (\ref{I0}). Examining $I_0$ near the point  
on the algebraic locus where $p^2=0$ and $m_1=m_2 = m$, to be referred to as $L1$, 
one finds that $\mu_0=\lam/(-4 p^2) \to m^2$ and \be
I_0 |_{L1} \simeq \frac{m}{\sqrt{-p^2}}\, j'(m)  ~.
\ee
Therefore $I_0 |_{L1}$ is divergent. Combining that with the finite value of $I_{L1}$ 
 (\ref{I_p2},\ref{I_lam})
we see that $C$ can be determined by requiring the boundary condition that $I$ is to be finite at $L1$.  In this sense \emph{the equation set suggests its own boundary conditions through regularity}. Evaluating (\ref{int1}) at L1, where $v \ll 1$ and hence $ B(v; a) \simeq v$ we conclude finally that  \be 
C=0 ~.
\label{bc}
\ee

We summarize that the result of integration and the application of boundary conditions is \be
I = \frac{\sqrt{\lam}}{-2 p^2} ~~ j'\(\frac{\lam}{-4 p^2} \) \[ B\(v_1; d/2-1 \) + B\(v_2; d/2-1  \) \] 
\label{soln}
\ee 
where $B(x;a)$ is defined in (\ref{def:B}) $j$ in (\ref{j_value}) and \be
 v_{1,2} := \frac{ \mp \Delta -p^2}{\sqrt{\lam}} 
\label{def:vi}
\ee
differ by signs only from $\tDelta_i$ (\ref{def:tDeltai})  and are chosen to make the expression manifestly symmetric with respect to $1 \leftrightarrow 2$ by using $B(-x;a)=-B(x;a)$.

\presub {\bf Comparison of the solution}. The integral (\ref{bubble-integ}) reduces to a single integration in alpha (or Schwinger) parameter space and can be evaluated directly, see appendix \ref{app:Schwinger} and in particular the expression (\ref{I_schw}). The solution (\ref{soln}) is equal to it.


The evaluation of the bubble integral appears in the literature: \cite{Davydychev2016} mentions e.g. \cite{BerendsDavydychevSmirnov1996,DavydychevDelbourgo1997} and the latter obtains the following expression (after accounting for different notation) \bea
I &=& i \pi^{d/2}\, \Gamma\(2-\frac{d}{2}\) \, \frac{1}{\sqrt{p^2}} \( \frac{m_1\, m_2\, \sin \tau_{12}}{\sqrt{p^2}}\)^{d-3} \[ \Omega_1 + \Omega_2 \] \non
\Omega_i &=& \int_0^{\tau_{0i}} \frac{d\theta}{\cos^{d-2} \theta}, ~~ i=1,2
\label{DavDelExp}
\eea
where the $\tau$ variables denote angles in the associated triangle. We confirmed that it equals (\ref{soln}) at least up to a phase depending on the analytic continuation. We note that the expression (\ref{soln}) has the minor advantage of recognizing the appearance of the tadpole integral $j'(\mu)$.

The result for the closely related two-loop vacuum diagram appeared in \cite{DavydychevTausk1992}.

\section{Triangle geometry}
\label{sec:triangle}

In this section we describe a geometrical interpretation of the variables which appear in the solution (\ref{soln}).

A hint is supplied by the definition of $\lam$ (\ref{def:lam}). It is a quadratic and symmetric function of its three arguments $\mu_1, \mu_2, p^2$. The symmetry suggest a possible relation with a triangle. Indeed, specifying edge lengths $a, b, c$ satisfying the triangle inequalities defines a unique triangle. Its area $A$ can be expressed in terms of the edge lengths, and is known since antiquity (at least as early as c. 60 AD) to be given by Heron's formula \cite{WikiHeron} \bea
 A^2 &=& s (s-a) (s-b) (s-c) = \non
 &=&\frac{1}{16} \( 2 a^2\, b^2 + 2 b^2\, c^2 + 2 a^2\, c^2 - a^4 - b^4 - c^4 \) = -\frac{1}{16} \lam \(a^2,b^2,c^2\)
 \label{Heron}
 \eea
 where \be
 s := \half \( a+ b+ c \) 
\ee
 is half the triangle circumference, and $\lam$, the K\"all\'en invariant, was defined in \eqref{def:lam}. In this sense the K\"all\'en invariant is essentially nothing more than the Heron formula, which we hence find to be the more appropriate name.

Indeed, it is known that this integral can be geometrically interpreted in terms of triangle geometry \cite{DavydychevDelbourgo1997,Davydychev2016} and references therein, where a geometrical interpretation was provided more generally to any 1-loop diagram, and in particular the relation between the K\"all\'en invariant and the area was noticed.

\begin{figure}
\centering \noindent
\includegraphics[width=8cm]{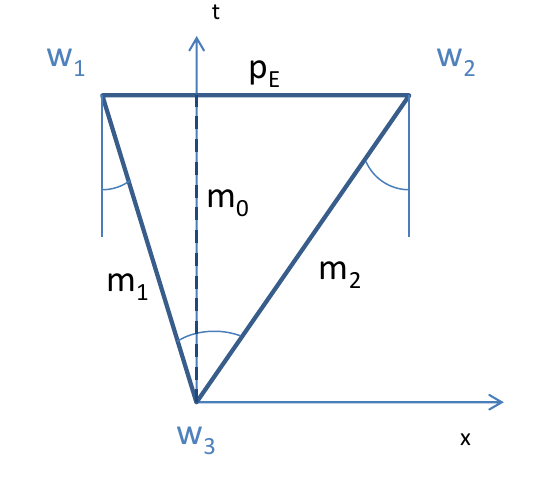}
\caption[]{The triangle corresponding to the parameters $\mu_1=m_1^2, \mu_2 = m_2^2, p^2=-p_E^2$. It is in the Minkowski plane (the plane has one time direction denoted by $t$ and one space direction denoted by $x$). The triangle defines 3 hyperbolic angles, namely rapidities, $w_1, w_2$ and $w_3$. $m_0$ is perpendicular to $p_E$.}
 \label{fig:triangle}
\end{figure}

Let us start by assuming the parameters are within the Euclidean domain (\ref{def:Euclid}). The parameters define a unique triangle in a plane with 1+1 signature, which has two time-like sides of lengths $m_1, m_2$ (since $\mu_1,\mu_2 \ge 0$) and whose third side is space-like and of length $p_E$ (since $p^2=-p_E^2<0$) as shown in fig. \ref{fig:triangle}. Note that while the parameters are in the Euclidean domain, the triangle is not in a Euclidean plane but rather in a Minkowski plane.

It is now suggestive to interpret the variables $v_{1,2}$ (\ref{def:vi}) in terms of the triangle. Since they are dimensionless it should be possible to express them in terms of the angles alone.  For triangles in the Euclidean plane it is well-known how to determine the angles in terms of the side lengths. For our triangle in the Minkowski plane it is not difficult to generalize these relations. The angle between the two time-like sides $m_1, m_2$ should be understood to be a hyperbolic angle, namely a rapidity $w_3$ which represents the required boost to make one of the sides parallel to the other. The expression for $w_3$ in terms of the side lengths is \be
 \cosh w_3 = \frac{\mu_1 + \mu_2 + p_E^2}{2 m_1 m_2}
\ee

The angle between a time-like and a space-like vectors is not Lorentz invariant. However, we can define a hyperbolic angle between one of the sides and the perpendicular to the other, as shown in the figure. These angles $w_i, i=1,2$ are given by \be
 \sinh w_i = \frac{p_E^2 \mp \Delta}{2 m_i\, p_E}
\ee
where $\Delta \equiv \mu_1-\mu_2$ and the minus sign corresponds to $w_1$. This implies \be
\cosh w_i = \frac{\sqrt{\lam}}{2 m_i \, p_E}
\ee
where in Minkowski space the Minkowski area $A_M$ requires changes the sign in  \eqref{Heron} into $A_M=\sqrt{\lam}/4$. Finally \be
 \tanh w_i  = \frac{\mp \Delta - p^2}{\sqrt{\lam}}
 \ee
 where $v_i$ denotes here the velocities associated with the rapidities $w_i$. These expressions are identical with those for the variables (\ref{def:vi}) appearing in the solution, and since the hyperbolic tangent of a rapidity is a velocity we obtain the geometrical interpretation for $v_{1,2}$ (\ref{def:vi}) , namely \be
\noindent\mbox{\begin{minipage}{\dimexpr\textwidth-2\fboxsep-2\fboxrule\relax}
\centering
$v_i$ is the velocity associated with the boost required to make the side $m_i$ perpendicular to the side $p_E$ 
\end{minipage}}
\label{interp}
\ee

The triangle geometry suggests also an interpretation for $\mu_0$ defined in (\ref{def:mu0}), namely 
\be
 \mu_0=m_0^2 
 \ee
 where $m_0$ is shown in fig. \ref{fig:triangle} and was already defined in \cite{DavydychevDelbourgo1997}. Indeed \be
 m_0 = \frac {2 A_M}{p_E} = \frac{\sqrt{\lam}}{2\sqrt{-p^2}} ~.
\ee
which implies (\ref{def:mu0}).

Comments \bi

\item For a Euclidean triangle the angles must sum up to $180^\circ$, namely $\al_1 + \al_2 + \al_3 = \pi$. Here the three rapidities are related by \be
 w_3 = w_1 + w_2
 \ee
 By a different sign conventions this can be brought to a form resembling the Euclidean one, namely $w_1 + w_2 + w_3 = 0$.

\item The analysis can be extended outside of the Euclidean domain. We find that \be
\noindent\mbox{\begin{minipage}{\dimexpr\textwidth-2\fboxsep-2\fboxrule\relax}
\centering
 Any set of parameters $\mu_1,\mu_2,p^2$ unconstrained by neither the triangle inequalities nor by positivity defines a signature for a plane and a triangle within it.
\end{minipage}}
\ee
All possible signatures are allowed, namely either $2+0,\, 1+1$ or $0+2$. The expressions of $v_i$ would be interpreted in terms of this triangle geometry. 

In more detail, there are three essential cases: if all the parameters are positive and they satisfy the triangle inequality then the triangle is within a two-time plane. If all are time like and $\sqrt{p^2} > m_1 + m_2$ then the signature is $1+1$ and $m_1$ and $m_2$ intersect in a point which is the future of the one and the past of the other. Finally if $m_1, m_2$ are time-like and either $p$ is space-like or $p$ is time-like and $p^2 < |m_1-m_2|$ then the signature is $1+1$ and $m_1, m_2$ intersect at a point which is either to the future of the two, or to the past of the two.

\ei

\section{Summary and discussion}
\label{sec:summ}

This paper studied the bubble diagram through the SFI method. While the result of the integral is known and is readily calculated in alpha parameters, this is the first diagram with external legs to be studied using SFI, and the original main goal had been to include external legs in the general method to determine the differential equation set. The main results of this paper are \bi

\item The original goal was achieved in (\ref{variation},\ref{preserve}). The relation with the vacuum closure of the diagram is given in (\ref{Gvac}).

\item The equations were solved in (\ref{soln}) reproducing the known full solution and hence demonstrating the consistency of SFI and moreover offering insight into ingredients of the general method of solving the SFI equation set, including transforming the equations into partially invariant variables. 

\ei

The correctness of the the SFI analysis of the bubble diagram was demonstrated by comparison to other methods, see the paragraph surrounding (\ref{DavDelExp}). The SFI treatment cannot be present already in the literature, as the method is recent. Finally, the main interest in the results is in the development of the SFI method, both in formulating the equations and in solving them, as it holds promise for novel evaluations of integrals which would result from this new conceptual framework. 

Additional results of this paper include (all referring specifically to the bubble diagram) \bi
\item The SFI equation set is given in (\ref{SFIbubble}), and in a $\gamma$ covariant basis in (\ref{cov_SFI_basis}).
\item The SFI group $G$ is given in (\ref{G}). Its orbits are found to be co-dimension 0, see ``Group orbits'' paragraph on p.8, 
thereby implying that the the SFI equation set determines the dependence of the integral on all of its parameters. 
\item \eqref{KObubble} states a relation between $I,\, \del I/\del(p^2)$ and the source $j$. Recalling that a bubble diagram is physically interpreted as self energy, this implies a relation between the correction to the mass, the field strength correction (or the renormalization of mass and field strength) and $j$ which holds in all dimensions.  
\item The equation set transformed into partially invariant variables is given in (\ref{transformed-set}).
\item The homogeneous solution is given in (\ref{I0}), the singular locus and the solution therein is given in (\ref{I_lam},\ref{I_p2}).
\item Following \cite{DavydychevDelbourgo1997} the triangle geometry (in non-Euclidean signatures) was discussed in section \ref{sec:triangle}. In particular a geometrical interpretation of the $v_{1,2}$ variables (\ref{def:vi}) appearing in the solution was given in (\ref{interp}).
\ei

\presub {Towards a general solution -- discussion}. We have seen that transforming into partially invariant variables appears to simplify the form of the equation set. This can possibly be quantified as follows. In general the SFI equation set allows us to express the Feynman integral as line integral within $G$-orbits in parameter space. Since $G$ orbits  of the bubble diagram are 3 dimensional, one can expect the line integral to consist of at most three integrals (along the various axes). However, the solution shows that a single integration suffices. Another tantalizing property of the solution is the decomposition into a sum of two terms.

We anticipate that a general method of solution should be possible to formulate, and such a method would specify the set of new variables, as well as the number of necessary integrations, and perhaps even the number of parameters present during each integration. Such a method is expected to reflect the structure of the differential equation set and in particular the group $G$. In this sense we anticipate that the structure of the group should be key to the solution method, just like Galois theory \cite{Galois} approaches the solution of algebraic equations through the structure of the associated discrete symmetry group.

\subsection*{Acknowledgments}

I would like to thank Philipp Burda, Ruth Shir and Erez Urbach for collaboration on related projects. 

This research is part of the Einstein Research Project ``Gravitation and High Energy Physics", funded by the Einstein Foundation Berlin, and it was also partly supported by the Israel Science Foundation grant no. 812/11 and the ``Quantum Universe'' I-CORE program of the Planning and Budgeting Committee.  

\appendix

\section{The induced SFI action}
\label{app:induced}

The action of $G$ on loop currents (\ref{def:vac-generator}) (in fact, a representation) induces a representation on both $S$, the space of squared propagator currents $k_i^2$, and on the space of parameters $\mu_i$ \be
T_C \to T_S,\, T_\mu ~.
\ee
Here we shall find relations between these representations. Here we shall consider a vacuum diagram. 

In order to discuss the action on $S$ we denote its basis by \be
 s_i := k_i^2 \equiv E_i^{jk} k_j\, k_k
 \label{def:si}
 \ee
 where the $E_i^{jk}$ tensor is defined by \be
 E_i^{jk} := \left\{ \begin{array}{cc}
  1 & \mbox{ for } i=j=k \\
  0 & \mbox{ otherwise}
  \end{array} \right.
  \label{def:Eijk}
 \ee
and it is introduced to avoid violation of the index summation convention. 
A generator $\(T_C \)_r^{~s}$ defined by (\ref{def:vac-generator}) acts on $s_i$ as follows \be
\delta s_i = 2 E_i^{jk}\, A_j^{~t}\, \( T_C\)_t^{~r}\, A_k^{~s} ~ l_r\, l_s ~,
\ee
where $A_i^{~r}$ is defined in (\ref{def:AirBiu}). By assumption $S$ is preserved by the action of $G$ and hence the same generator can be represented also by \be
 \delta s_i = \(T_S\){i}^{~~j}\, s_j ~.
\label{var-si}
\ee
Expanding the previous expression \be
  \delta s_i  = \(T_S\)_i^{~~j}\, E_j^{kl}\, A_k^{~r}\, A_l^{~s} ~ ~ l_r\, l_s ~.
 \ee
 Hence the two representations $T_r^{~s}$ and $T_i^{~j}$ are related by \be
  E_i^{jk}\, A_j^{~t}\, \(T_C\)_t^{~(r}\, A_k^{~s)} = \(T_S\)_i^{~~j}\, E_j^{kl}\, A_k^{~r}\, A_l^{~s} 
 \ee

The action on the $\mu$ parameter space can be obtained by the following consideration. The integrand $\tI$ of a Feynman diagram is a function of the propagator denominators 
 $E_i := s_i - \mu_i$ namely \be
 \tI = \tI\( E_i \) \equiv \tI \( s_i -\mu_s\) ~.
 \ee
Hence the variation of $s_i$ (\ref{var-si}) induces the following variation of $\tI$ \be
	\delta \tI = \frac{\del \tI}{\del E_i}\, \delta s_i = \frac{\del \tI}{\del E_i}\, T_{S\,i}^{~~j}\, s_j  
	\ee
Using \bea
 \frac{\del}{\del E_i} &=& - \frac{\del}{\del \mu_i}	 \non 
 			s_j	&\to& \mu_j
 \eea
where the last replacement neglects $E_j$ terms which are grouped under the source terms, we obtain
\bea
 \delta \tI &=& O\,  \tI \non 
 O &:=&  - \( T_S\)_i^{~j} \mu_j\, \frac{\del}{\del \mu_i}\
 \eea
Hence we finally define \be
 T_\mu = -\(T_S\)^T
 \ee
 where the transpose operation is inserted so that commutators of the operators $O$ get mapped unto commutators of the matrices $T_\mu$. $T_\mu$ is a representation of $G$ known as the \emph{dual representation} to $T_S$. 
  
 We comment on a geometrical interpretation of $\mu$ space. $s_i$ is a basis for the space $S$, hence a general element can we expressed as \be
  s = s_i \alpha^i
  \ee
  where the coordinates $\alpha^i$ can be interpreted as Schwinger parameters. The $\mu_i$ are dual to $\alpha^i$ (in fact related by a Legendre transform, see e.g. \cite{locus}) hence \emph{$\mu_i$  are coordinates on the dual space $S^*$}.

\section{Integration in alpha parameter space} 
\label{app:Schwinger}


The bubble diagram can be readily evaluated in the space of alpha (or Schwinger) parameters.\footnote{
The alpha parameters are already mentioned in the appendix of Feynman's first paper on his diagrams \cite{FeynDiag}, where it says after eq. (14a)``suggested by some work of Schwinger's involving Gaussian integrals''.} 
For easy reference, we include the computation here. We follow the conventions of \cite{SmirnovBooks}.


We start with the tadpole diagram \bea
  j(\mu) &\equiv& I_{fig. \ref{fig:bubble}(b)} =  -\int d^dl\, \frac{dl}{-l^2 + \mu-i 0}= -i \, \int dl_E\, \frac{dl}{l_E^2 + \mu} = -i \,  \int dl_E\, d\alpha\, \exp \(-\alpha (l_E^2 + \mu) \) =  \non
  		&=&  -i \,  \int_0^\infty d\alpha\, \exp (-\alpha\, \mu )\, \( \frac{\pi}{\alpha} \)^{d/2} = 
		-i \pi^{d/2}\, \Gamma\(1-\frac{d}{2}\)\, \mu^{\frac{d}{2}-1}
 \label{j-computed}
  \eea
where we used the standard substitution $l_0\ to i\, l_{E0}$. \eqref{j-computed} implies  \be
 j '(\mu) =  i \pi^{d/2}\, \Gamma\(2-\frac{d}{2}\)\, \mu^{\frac{d}{2}-2}
 \ee
 Note that as usual factors of $i \pi^{d/2}$ can be eliminated from the results if we adopt the conventions that each loop integral is to be normalized as follows $-i \int d^dl/\pi^{d/2}$.
 
The bubble diagram evaluates to \bea
I(p^2 \equiv -p_E^2;\, \mu_1,\mu_2 ) &=&  I_{fig. \ref{fig:bubble}(a)} = \int \frac{d^d l} { \( -(p/2+l)^2+\mu_1 \)\,   \( -(p/2-l)^2+\mu_2 \)} = \non
		&=& i \int \frac{dl_E} { \( (p_E/2+l_E)^2+\mu_1 \)\,   \( (p_E/2-l_E)^2+\mu_2 \)} \non
 	&=& i\, \int dl_E d\al_1 d\al_2\, \exp \[- \( \al_1 \[ \(p_E/2+l_E\)^2 + \mu_1 \]+ \al_2 \[ \(p_E/2-l_E\)^2 + \mu_2 \] \) \] \non
	&=&  i\, \int d\al_1 d\al_2\, \( \frac{\pi}{\al_1 + \al_2} \)^{d/2} \exp \[ - \( \al_1 \mu_1 + \al_2  \mu_2 + \frac{\al_1 \al_2}{\al_1 + \al_2} p_E^2 \) \] = \non
	&=& i \pi^{d/2} \int \al d\al\, d\beta_1 d\beta_2\, \delta\(\beta_1 + \beta_2 -1\)\, \frac{1}{\al^{d/2}}\, \exp \[ -\al \( \beta_1 \mu_1 + \beta_2  \mu_2 + \beta_1 \beta_2\, p_E^2 \) \] = \non 
	&=& i \pi^{d/2}\, \int_0^1 d\beta\,  \Gamma\( 2-\frac{d}{2}\) \[ \beta \mu_1 + (1-\beta)  \mu_2 + \beta (1-\beta)\, p_E^2 \] ^{d/2-2}  = \non
	&=& i \pi^{d/2}\,\Gamma\( 2-\frac{d}{2}\)\, \frac{\sqrt{\lam}}{2 p_E^2}\, \( \frac{\lam}{4 p_E^2} \)^{\frac{d}{2}-2}  
	 \int_{\tbt_1}^{\tbt_2} \(1-\tbt^2 \)^{d/2-2} = \non
	&=& i \pi^{d/2}\,\Gamma\( 2-\frac{d}{2}\)\, \frac{\sqrt{\lam}}{2 p_E^2}\, \( \frac{\lam}{4 p_E^2} \)^{\frac{d}{2}-2}  \[B\(\tbt_+; d/2-1\) - B\(\tbt_-; d/2-1\) \]  = \non 
	&=& i \pi^{d/2}\,\Gamma\( 2-\frac{d}{2}\)\, \frac{\sqrt{\lam}}{2 p_E^2}\, \( \frac{\lam}{4 p_E^2} \)^{\frac{d}{2}-2}  \[B\(\tbt_+; d/2-1\) + B\( -\tbt_-; d/2-1\) \] \non
\label{I_schw}
\eea
where in passing to line 7 the integration variable was changed according to \be
 \tbt := \frac{2\bt - \bt_+ - \bt_-}{\bt_+ - \bt_-}
 \label{def:tbt}
 \ee
 where \bea
  \bt_\pm &:=& \frac{1}{2 p_E^2} \[ \mu_1-\mu_2 + p_E^2 \pm \sqrt{\lam} \] \non
  \lam &=& \lam (\mu_1,\mu_2,p^2) \equiv \lam (\mu_1,\mu_2, -p_E^2)  
\eea
and the new integration limits are \be
\tbt_\pm := \frac{1}{\sqrt{\lam}}\, \( -\mu_1 + \mu_2  \mp p^2 \) \equiv \frac{1}{\sqrt{\lam}}\, \( -\mu_1 + \mu_2  \pm p_E^2 \) 
\label{def:tbtpm}
\ee

We note that in the Euclidean domain (\ref{def:Euclid})  \be
 \lam \ge 0, \qquad -1 \le \tbt_- \le \tbt_+ \le 1
 \ee
 (assuming $p_E^2,\mu_1,\mu_2 \ge 0$) so that the arguments of $B(x;a)$ within its natural range and are univalued. In Lorentzian signature an analytic continuation of $B$ may be required.

\section{Comparison with standard IBP approach} 

The bubble diagram is simple enough that it can be evaluated directly through alpha parameters. Still, it is interesting to compare the SFI treatment of the bubble diagram in this paper, with more standard treatments (namely, non SFI) in the literature.

In the standard IBP treatment one fixes the masses, and seeks recursion relations for the propagator indices which determine the master integrals, see e.g. \cite{SmirnovBooks}, possibly with the help of a computer program such as FIRE \cite{FIRE}. 

Especially interesting are the cases when an integral with unit indices can be expressed as a sum of simpler integrals. This happens when the list of master integrals does not include the original integral, and in this sense the master integrals are non-trivial. From the SFI perspective this means that the parameters are within the algebraic locus. In this paper the algebraic locus was found to consist of 
two components: the $\lam=0$ cone and the $p^2=0$ plane (\ref{sing}). 
I note that FIRE requires the user to supply a list of a subset of the IBP relations to be used, but this is not required in the SFI approach. 

For generic values of the parameters I would expect the list of master integrals to include the original integral, and in that sense, one would be compelled to use other methods, such as integration of alpha parameters or the method of Differential Equations. Similarly, the SFI method generates a set of 3 partial differential equations, which were used here to determine the solution.

All methods find that the integral can be expressed as \emph{a sum of two terms}, and moreover \emph{each term contains a single integral}. \cite{DavydychevDelbourgo1997} presented an attractive geometrical interpretation for this decomposition in terms of triangle geometry. Given a general diagram, it would be very interesting to be able to anticipate whether a similar decomposition exists, how many terms it would have, and how many integrations would be required. So far the SFI approach did not shed light on these questions, but it might.

Finally I mention that the The SFI approach recognizes  a term in the solution (\ref{soln}) as the tadpole diagram (\ref{j_value}). This can be traced to the appearance of the tadpole diagram as a source in the SFI equation set (\ref{SFIbubble}).

\newpage
\bibliographystyle{unsrt}

\end{document}